
\magnification=\magstep1

\hsize13.3cm
\vsize20.cm
\line{\hfill HD-THEP-94-21}
\vskip 1cm
\noindent
\centerline{\bf HIGH TEMPERATURE FIELD THEORY}
\centerline{\bf BEYOND PERTURBATION THEORY}
\vskip 1cm
\centerline{Christof Wetterich}
\bigskip
\centerline{Institut  f\"ur Theoretische Physik}

\centerline{Universit\"at Heidelberg}

\centerline{Philosophenweg 16}

\centerline{D-6900 Heidelberg, FRG}
\vskip 1.5cm\noindent
{\bf ABSTRACT}
\bigskip
For realistic values of the Higgs boson mass the high temperature
electroweak phase transition cannot be described perturbatively.
The symmetric phase is governed by a strongly interacting $SU(2)$ gauge
theory. Typical masses of excitations and scales of condensates are set
by the ``high temperature confinement scale'' $\approx 0.2\ T$.
For a Higgs boson mass around 100 GeV or above all aspects of the phase
transition are highly nonperturbative. Near the critical temperature
strong electroweak interactions are a dominant feature also in the
phase with spontaneous symmetry breaking. Depending on the value of the
Higgs boson mass the transition may be a first order phase transition
or an analytical crossover.

\vskip1cm\noindent
{\bf 1. RUNNING COUPLINGS NEAR PHASE TRANSITIONS}
\bigskip
Phase transitions in the very early universe may be tumultuous
periods out of thermodynamic equilibrium. Some traces of such a
transition may still be observable today, giving a hint on what happend
shortly after the big bang. Remnants of the electroweak phase transition
may help to understand the universe at the age of about $10^{-12}s$. One
of the most prominent features of the standard model of electroweak
interactions is spontaneous symmetry breaking. The masses of the gauge
bosons and fermions are proportional to the vacuum expectation value of
the Higgs doublet. According to the original argument of
Kirzhnits and Linde$^1$
this expectation value vanishes at sufficiently high temperature and
the electroweak symmetry is restored. Such high temperatures were
presumably realized in the very early universe immediately after the
big bang. As the universe cooled there must have been a phase transition
from the the symmetric to the spontaneously broken phase of the standard
model. This phase transition may have important consequences for our
present universe, one example being the possible creation of the excess
of matter compared to antimatter (baryon asymmetry)$^2$. The physical
implications of the high temperature electroweak phase transition depend
strongly on its nature (whether it is second or first order) and its
details. More specifically, the possibility of generating the baryon
asymmetry imposes the requirement of out of equilibrium conditions$^3$,
which is satisfied only if the phase transition is of the first order;
the exact amount of produced baryon number is very sensitive to the
details
of the fluctuations which drive the transition (profile of bubbles,
velocity
of the wall etc.)$^4$; and avoiding the washing out of any generated
baryon
asymmetry requires a sufficiently strong first order phase transition$^5$.

Most theoretical studies of the electroweak phase transition use high
temperature perturbation theory$^6$ for a computation of the
temperature-dependent effective potential for the Higgs field. This is
often
supplemented by an appropriate resummation of graphs or a solution of a
corresponding gap equation for the mass terms$^7$. Near a phase
transition,
however, the use of perturbation theory becomes questionable. Roughly
speaking, one can trust perturbation theory only to the extent that mean
field theory gives a qualitatively correct description of the
corresponding
statistical system. This is known to be not always the case for critical
phenomena. A good example for the breakdown of high temperature
perturbation
theory in the vicinity of the phase transition are scalar field
theories$^8$.
The second order character of the transition and the corresponding
critical
exponents are not reproduced by perturbation theory.

The deeper reason for the breakdown of perturbation theory lies in the
effective three-dimensional character of the high temperature
field theory$^9$.
Field theory at nonvanishing temperature $T$ can be formulated in
terms of an Euclidean functional integral where the ``time dimension''
is compactified on a torus with radius $T^{-1}$ $^{10}$. For phenomena
at distances larger than $T^{-1}$ the Euclidean time dimension cannot
be resolved. Integrating over modes with momenta $p^2>(2\pi T)^2$ or,
alternatively, over the higher Fourier modes on the torus (the $n\not=0$
Matsubara frequencies) leads to ``dimensional reduction'' to an effective
three-dimensional theory. This is very similar to dimensional reduction in
Kaluza-Klein theories$^{11}$ for gravity. The change of the effective
dimensionality for distances larger than $T^{-1}$ is manifest
in the computation$^8$ of the temperature-dependent effective potential
in scalar theories. The scale dependence of the renormalized couplings
is governed by the usual perturbative $\beta$-functions only for
$p^2>(2\pi T)^2$.
In contrast, for smaller momenta $p^2<(2\pi T)^2$ the running of the
couplings was found to be determined by three-dimensional
$\beta$-functions
instead of the perturbative four-dimensional ones - as proposed in a
different setting in ref. 12. The effect of the three-dimensional running
is
clearly manifest in the temperature-dependence of the couplings shown$^8$
in fig. 1, especially for the renormalized quartic scalar coupling
$\lambda_R$, which vanishes for $T$ approaching the critical temperature
$T_c$.
As an alternative to integrating out all modes with $p^2>(2\pi T)^2$ an
effective three-dimensional theory for the long distance electroweak
physics
has been obtained in ref. 13 by integrating out the higher Matsubara
frequencies\footnote*{For an earlier treatment of dimensional reduction
in high temperature QCD see ref. 14.}.

If the three-dimensional running of the couplings becomes important, the
physics of the phase transition is dominated by classical statistics even
in case of a quantum
field theory. A second order phase transition is characterized
by an infinite correlation length. The critical exponents which describe
the behaviour near the critical temperature are always those of the
corresponding classical statistical system. Since the fixpoints of the
three-dimensional $\beta$-functions are very different from the
four-dimensional (perturbative) fixpoints, we conclude that high
temperature
perturbation theory is completely misleading in the vicinity of a
second-order phase transition. This argument extends to sufficiently weak
first order transitions. A second related example for the breakdown
of perturbation theory is the symmetric phase of the electroweak gauge
theory. The gauge bosons are massless in perturbation theory and the
three-dimensional running always dominates at large distances$^{15}$.

In order to understand the high temperature behaviour of a theory
we should understand the qualitative features of the $\beta$-functions
in three dimensions. These $\beta$-functions have nothing to do with the
ultraviolet regularization of the field theory - in this respect there is
no difference between vanishing and nonvanishing temperature. They are
rather related to the infrared behaviour of the theory or the dependence
of Green functions on some sort of infrared cutoff. According to Wilson's
concept of the renormalization group these $\beta$-functions describe the
scale
dependence of the couplings if one looks at the system on larger and
larger
distances. For an understanding of systems with approximate scaling in a
certain range it is useful to define dimensionless couplings. One divides
out an appropriate power of the infrared cutoff $k$ which plays the role
of the renormalization scale. For example, the dimensionless quartic
scalar
coupling $\lambda$ in the effective three-dimensional theory is related
to the four-dimensional coupling $\lambda_4$ and the temperature by
$$\lambda={\bar\lambda_3 \over k}=\lambda_4{T \over k}\eqno(1)$$
For a pure $N$-component scalar theory the qualitative behaviour of the
function $\beta_\lambda={\partial\lambda \over \partial t}, \
t=\ln k$, is shown in fig. 2. Arrows denote the flow of $\lambda$ with
decreasing $k$ and we have assumed a massless theory. One clearly
sees an infrared stable fixpoint corresponding to the second-order phase
transition.

Next we turn to scalar QED which should describe the phase transition
of superconductors or of the abelian Higgs model at high temperatures.
The supposed form of the functions $\beta_{e^2}=\partial e^2/\partial t$
and $\beta_{\lambda/e^2}=\partial\left({\lambda \over e^2}\right)/\partial
t$ is shown\footnote*{The true phase diagrams are multidimensional and the
diagrams of fig. 3 may be interpreted as projections on appropriate
trajectories. The fixpoint structure is independent of particular
trajectories.}
in fig. 3. One observes a fixpoint for $e^2$ but no fixpoint for
$\lambda/e^2$
or $\lambda$. Therefore $\lambda$ decreases until it vanishes and the
corresponding phase transition is presumably of the first order. For
comparison we also show the $\beta$-function for a number $M$ of complex
scalar fields exceeding a critical value $M_{cr}$. Here an ultraviolet
and an infrared stable fixpoint appear. For small initial $\lambda$
($\lambda/e^2$ smaller than the UV fixpoint) the phase transition remains
of the first order. The UV fixpoint corresponds to a triple point and for
initial $\lambda/e^2$ larger than the UV fixpoint one has a second-order
transition with critical behaviour governed by the IR fixpoint. The
cirtical
number of charged scalars $M_{cr}$ (above which a second order transition
is possible) is not very well known.

A nonabelian gauge theory like the electroweak theory is confining also in
three dimensions. We have depicted the running of the gauge coupling $g^2$
and
the ratio $\lambda/g^2$ in fig. 4. For sufficiently small initial
$\lambda$
(small physical Higgs boson mass) $\lambda(k)$ reaches zero for $k$ much
larger
than the three-dimensional confinement scale. One then expects a
first-order
 transition which is analogous to the four-dimensional Coleman-Weinberg
scenario$^{16}$. Typical mass scales are of the order $k_{cw}$ where
$\lambda(k_{cw})=0$. In this case  it is possible that high temperature
perturbation theory gives reliable results. On the other hand, if the
three-dimensional confinement scale $\Lambda_{conf}^{(3)}$ (the value
of $k$ for which the gauge coupling diverges or becomes very large) is
reached with $\lambda(\Lambda^{(3)}_{conf})>0$ the behaviour near the
phase transition is described by a strongly interacting electroweak
theory.
Then strong effective coupling constants appear not only in the symmetric
phase, but also in the phase with spontaneous symmetry breaking. The phase
transition may either be of the first order or an analytical crossover
may replace the phase transition$^{15}$. A second-order transition
seems unlikely.\footnote*{A second-order transition requires a massless
scalar degree of freedom at the critical temperature. The Higgs scalar
presumably acquires a mass through strong electroweak interactions
both in the symmetric and SSB phase. Chiral condensation phenomena
of quarks and leptons similar as for quarks in QCD seem not very plausible
for the high temperature electroweak theory.} In any case, the
``strongly interacting phase transition'' will be very different from
perturbative expectations.

\vskip1cm\noindent
{\bf 2. PERTURBATIVE INFRARED DIVERGENCES AND ``CUBIC TERMS''}
\bigskip
In high temperature perturbation theory the electroweak phase transition
is predicted to be of the first order. This follows from a
``cubic term'' in the effective potential generated by thermal
fluctuations.
In order to get an idea for which values of the scalar field and the
temperature
perturbation theory may be applied, we have to estimate the reliability
of the cubic term. For a complex two-component Higgs scalar $\varphi$ the
cubic term is proportional $(\varphi^\dagger\varphi)^{3/2}$ and we should
understand
the origin of such a nonanalytic behaviour of the effective potential. We
will see that it is closely linked to the issue of infrared divergences
for the quartic scalar coupling$^8$. Indeed, a term
$\sim(\varphi^\dagger\varphi)
^{3/2}$ implies that the quartic scalar coupling diverges for
$\varphi\to0$. This infrared divergence$^{17}$ is immediately apparent by
inspection of the
diagram in fig. 5 for which the $n=0$ Matsubara frequency gives a
contribution
$$\Delta\lambda_4\sim T\int d^3q{g_4^4 \over (q^2+{1 \over
2}g_4^2\varphi^\dagger\varphi)^2}\sim
{g_4^3T \over (\varphi^\dagger\varphi)^{1/2}}\eqno(2)$$
The $\varphi$-dependent gauge boson mass plays here the role of the
effective infrared cutoff and we may associate
$$k=m_W(\varphi)={1 \over \sqrt2}
g_4(\varphi^\dagger\varphi)^{1/2}\eqno(3)$$
Inserting the correction to the quartic term $\sim\Delta
\lambda\ (\varphi^\dagger\varphi)^2$
in the effective potential gives exactly the cubic term mentioned before.
The
question arises now to which extent the perturbative treatment of the
effectively three-dimensional behaviour can be trusted. A similar cubic
term appears also in the perturbative treatment of the pure scalar theory
and has been shown$^8$  to be completely misleading in the vicinity of
the critical temperature. (For $T=T_c$ the effective scalar potential
has not a divergent, but a vanishing quartic coupling and is dominated
by the six-point function, $U\sim(\varphi^\dagger\varphi)^3$.) What about
gauge theories?

At the scale $k_T=2\pi T$ the three-dimensional gauge coupling and quartic
scalar coupling are given by
$$\bar\lambda_3(k_T)=\lambda_4T,\quad\bar g^2_3(k_T)=g^2_4T\eqno(4)$$
For $k<k_T$ the running of the couplings becomes effectively
three-dimensional.
In the limit of small $\bar\lambda_3\ll\bar g^2_3$ we neglect the
contributions
from scalar loops and the running of $\bar\lambda_3$ follows\footnote*{If
the
contribution of the Debye-screened $A_0$ mode is omitted, the r.h.s. of
eq. (5) should be reduced by a factor 2/3, see sect. 5.} from gauge boson
loops (fig. 5)
$${\partial\bar\lambda_3 \over \partial t}={9 \over 64\pi} {\bar g^4_3
\over k}
\eqno(5)$$
Let us for a moment neglect the running of the gauge coupling. With the
infrared cutoff identified with $m_W(\varphi)$ (3) and defining
$\bar\lambda_3$
by an appropriate derivative of the effective potential $U$
$$\eqalign{
\bar\lambda_3(k)&=T{\partial^2U \over \partial\rho^2}\cr
\rho&=\varphi^\dagger\varphi=\varphi^2,\quad k^2={1 \over 2}
g^2_4\rho\cr}\eqno(6)$$
the flow equation (5) can be turned into a differential equation for
$U(\rho)$. Solving
$$\varphi{\partial \over \partial\varphi} \bar\lambda_3={9\sqrt2 \over
64\pi}
g^3_4{T^2 \over \varphi}\eqno(7)$$
for $\varphi<\varphi_T,{1 \over 2} g^2_4\varphi^2_T=k^2_T=(2\pi T)^2$ one
obtains
$$\bar\lambda_3(\varphi)=\left(\lambda_4+{9g^4_4 \over 128\pi^2}\right)
T-{9\sqrt2g^3_4 \over 64\pi} {T^2 \over \varphi}\eqno(8)$$
{}From the definition (6) of $\bar\lambda_3$ follows the second order
differential equation
$${\partial^2 U \over \partial \rho^2}=\lambda_4+{9 g^4_4 \over
128\pi^2}-{9\sqrt 2 g^3_4 \over 64\pi}T\rho^{-{1\over2}}\eqno(9)$$
The general solution has two integration constants, one of them being
irrelevant here
$$U(\rho)=-\mu^2(T)\rho+{1 \over 2}(\lambda_4+{9g^4_4 \over
128\pi^2})\rho^2-
{3\sqrt2 \over 16\pi} g^3_4 T\rho^{3/2}\eqno(10)$$
The temperature-dependent mass term $\mu^2(T)$
appears here as an integration constant and can be determined
from high temperature perturbation theory (which should be valid for
$\varphi=\varphi_T$)
$$-\mu^2(T)=-\mu^2_0+{3g^2_4 \over 16}T^2\eqno(11)$$
The potential (10) is nothing else than the result of high temperature
perturbation theory. We conclude that the latter is reliable if the scalar
loops can be omitted
$(\bar\lambda_3\ll \bar g^2_3)$ and if the running of the gauge coupling
can be neglected. Additional terms from the scalar loops could be
included in the evolution equation (5) without changing the qualitative
picture. The crucial question concerns the neglection of the running
of the gauge coupling. This will determine the range of validity of
perturbation
theory and we will turn back to this question below.

\vskip1cm\noindent
{\bf 3. AVERAGE ACTION}
\bigskip
A useful tool for describing the running of couplings in arbitrary
dimension is the average action$^{18}$. Consider a simple model with a
real scalar field
$\chi$. The average scalar field is easily defined by
$$\phi_k(x)=\int d^dyf_k(x-y)\chi(y)\eqno(12)$$
with $f_k$ decreasing rapidly for $(x-y)^2>k^{-2}$ and properly
normalized. The average is taken over a volume of size $\sim k^{-d}$.
The average action $\Gamma_k[\varphi]$ obtains then by functional
integration of the
``microscopic variables'' $\chi$ with a constraint forcing $\phi_k(x)$ to
equal
$\varphi(x)$ up to small fluctuations. It is the effective action for
averages
of fields and therefore the analogue in continuous space of the block spin
action$^{19}$ on the lattice. All modes with momenta $q^2>k^2$ are
effectively
integrated out. Lowering $k$ permits to explore the theory at longer and
longer distances. The average action has the same symmetries as the
original action. As usual it may be expanded in derivatives, with average
potential $U_k(\rho),\rho={1 \over 2}\varphi^2$, kinetic term, etc.
$$\Gamma_k=\int d^dx\left\lbrace U_k(\rho)+{1 \over 2}
Z_k(\rho)\partial_\mu\varphi\partial^\mu\varphi+...
\right\rbrace\eqno(13)$$

In a suitable formulation$^{20}$ the effective average action becomes the
generating functional for 1PI Green functions with an infrared cutoff set
by
the scale $k$. It interpolates between the classical action for
$k\to\infty$ and the effective action for $k\to0$. In this version an
exact nonperturbative
evolution equation describes the dependence of $\Gamma_k$ on the infrared
cutoff $k$ $(t=\ln k)$
$${\partial \over \partial t}\Gamma_k={1 \over 2} \int {d^dq \over
(2\pi)^d}
\left(\Gamma_k^{(2)}+R_k\right)^{-1}{\partial \over \partial
t}R_k\eqno(14)$$
Here $R_k(q)$ is a suitable infrared cutoff which may depend on $q^2$, as,
for
example,\hfill\break
$R_k=q^2\exp\left(-{q^2 \over k^2}\right)\left(1-\exp\left(-{q^2 \over
k^2}\right)\right)^{-1}$ or $R_k=k^2$. The
two-point function $\Gamma_k^{(2)}$ obtains by second functional variation
of $\Gamma_k$
$$\Gamma_k^{(2)}(q',q)={\delta^2\Gamma_k \over
\delta\varphi(-q')\delta\varphi(q)}\eqno(15)$$
Therefore $(\Gamma_k^{(2)}+R_k)^{-1}$ is the exact propagator in presence
of
the infrared cutoff $R_k$ and the flow equation (14) takes the form of
the scale variation of a renormalization group-improved one-loop
expression. We emphasize that the evolution equation is fully
nonperturbative and no approximations are made. A simple proof can be
found
in ref. 20. The exact flow equation (14) can be shown to be equivalent
with earlier versions of ``exact renormalization group equations''$^{21}$
and it encodes the same information as the Schwinger-Dyson
equations$^{22}$.

An exact nonperturbative evolution equation is not yet sufficient for an
investigation of nonperturbative problems like high temperature field
theories. It is far too complicated to be solved exactly. For
practical use it is crucial to have a formulation that allows to find
reliable nonperturbative approximative solutions. Otherwise speaking,
one needs a description of $\Gamma_k$ in terms of only a few $k$-dependent
couplings. The flow equations for these couplings can then be solved
numerically or by analytical techniques. It is on the level of such
truncations of the effective average action that suitable approximations
have to be found. In this respect the formulation of the effective
average action$^{20}$ offers important advantages: The average action has
a simple physical interpretation and eq. (14)
is close to perturbation theory if the
couplings are small. The formulation is in continuous space and all
symmetries - including chiral symmetries or gauge symmetries$^{15}$  - can
be respected. Since $\Gamma_k$
has a representation as a functional integral alternative methods
(different
from solutions of the flow equations) can be used for an estimate of its
form. Furthermore, the flow equation (14)
is directly sensitive to the relevant infrared physics
since the contribution of particles with mass larger than $k$ is
suppressed by the propagator on the r.h.s. of eq. (14).  The closed form
of this equation does not restrict one a priori to
given expansions, like in 1PI $n$-point functions.
In addition the momentum integrals in eq. (14) are both infrared
and ultraviolet convergent if a suitable cutoff $R_k$ is chosen. Only
modes in the vicinity of $q^2=k^2$ contribute substantially. This
feature is crucial for gauge theories where the formulation of a
gauge-invariant ultraviolet cutoff is difficult without dimensional
regularization.

\vskip1cm\noindent
{\bf 4. STRONG ELECTROWEAK INTERACTIONS}
\bigskip
We are now ready to discuss the running of the three-dimensional
gauge coupling. We start from the effective average action for a pure
$SU(N_c)$
Yang-Mills theory. It is a gauge-invariant functional of the gauge field A
and obeys the exact evolution equation$^{15}$  (with Tr including a
momentum
integration)
$${\partial \over \partial t}\Gamma_k[A]={1 \over 2} {\rm Tr}
\left\lbrace{\partial R_k[A] \over \partial
t}\left(\Gamma_k^{(2)}[A]+GF[A]+R_k[A]\right)^{-1}
\right\rbrace-\epsilon_k[A]
\eqno(16)$$
Here $GF[A]$ is the contribution from a generalized gauge-fixing term in a
covariant background gauge
$$GF[A]=\Gamma_k^{gauge(2)}[A,\bar A]_{|\bar A=A}\eqno(17)$$
and $\epsilon_k[A]$ is the ghost contribution$^{15}$. The infrared cutoff
$R_k$ is in general formulated in terms of covariant derivatives. We make
the
simple truncation
$$\eqalign{
\Gamma_k[A]&={1 \over 4}\int d^dxZ_{F,k}F_{\mu\nu}F^{\mu\nu}\cr
\Gamma_k^{gauge}[A,\bar A]&={1 \over 2}\int d^dxZ_{F,k}(D_\mu[\bar
A](A^\mu-
\bar A^\mu))^2\cr}\eqno(18)$$
and observe that in our formulation the gauge coupling $\hat g$ appearing
in $F_{\mu\nu}$ and $D_\mu$ is a constant independent of $k$. The only
$k$-dependent coupling can be associated with the dimensionless
renormalized
gauge coupling
$$g^2(k)=k^{d-4}\bar g^2_d(k)=k^{d-4}Z^{-1}_{F,k}\hat g^2.\eqno(19)$$
The running of $g^2$ is related to the anomalous dimension $\eta_F$
$$\eqalign{
\eta_F&=-{\partial \over \partial t}\ln Z_{F,k}\cr
{\partial g^2 \over \partial t}&=\beta_{g^2}=(d-4)g^2+\eta_F
g^2\cr}\eqno(20)$$
Using configurations with constant magnetic field it was found$^{15}$ to
obey
$${\partial g^2 \over \partial t}=(d-4)g^2-{{44 \over 3}N_cv_da_dg^4 \over
1-{20 \over 3} N_c v_d b_dg^2}\eqno(21)$$
with\footnote*{The vanishing of the $\beta$-function for $\bar g^2$ for
$d=2$ and $d=26$ is no accident. In lowest order in the
$\epsilon$-expansion$^{23}$
the denominator in the last term is absent and $v_3a_3$ is replaced by
$v_4a_4$.}
$$\eqalign{
v^{-1}_d&=2^{d+1}\pi^{d \over 2}\Gamma\left({d \over 2}\right)\cr
a_d&={(26-d)(d-2) \over 44}n^{d-4}_1\cr
b_d&={(24-d)(d-2) \over 40}l^{d-2}_1\cr}\eqno(22)$$
Only the momentum integrals $(x\equiv q^2)$
$$\eqalign{
n^d_1&=-{1 \over 2}k^{-d}\int^\infty_0 dx\ x^{d \over 2} {\partial \over
\partial
t}\left({\partial P \over \partial x}P^{-1}\right)\cr
l^d_1&=-{1 \over 2}k^{2-d}\int^\infty_0 dx\ x^{{d \over 2}-1} {\partial
\over \partial
t} P^{-1}\cr}\eqno(23)$$
depend on the precise form of the infrared cutoff $R_k$ appearing in
$$P(x)=x+Z^{-1}_kR_k(x)\eqno(24)$$
In four dimensions one has $a_4=1,\ v_4=1/32\pi^2$ and eq. (21) reproduces
the one-loop result for $\beta_{g^2}$ in lowest order  $g^4$. For an
appropriate choice of $R_k$ with $b_4=1$ an expansion of eq. (21)
also gives 93 \% of the perturbative $g^6$ coefficient. For $d\not=4$ the
solution of eq. (21) is
$$\eqalign{
&{g^2(k) \over (1+\delta g^2(k))^\gamma}={g^2(k_0) \over (1+\delta
g^2(k_0))^\gamma}\left({k \over k_0}\right)^{d-4}\cr
&\delta={N_c v_d \over 3}\left({44 \over 4-d} a_d-20b_d\right)\cr
&{1 \over \gamma}=1-{5 \over 11}(4-d){b_d \over a_d}\cr}\eqno(25)$$

Concerning the high temperature field theory we should use the
three-dimensional $\beta$-function and associate $k_0$ with the scale
$k_T=2\pi T$, where the three-dimensional running sets in. The ``initial
value'' of the
gauge coupling reads $g^2(k_T)=2\alpha(k_T)$ with $\alpha$ the
four-dimensional fine structure constant. For $k<k_T$ the
three-dimensional
gauge coupling increases with a power behaviour instead of the
four-dimensional
logarithmic behaviour. The three-dimensional confinement scale
$\Lambda_{conf}
^{(3)}$ - where $g^2$ diverges - is proportional to the temperature. For
the
electroweak theory $(\alpha_w=1/30)$ and the choice (24) $P(x)=x/(1-\exp-
{x \over k^2})$ one finds$^{15}$
$$\Lambda^{(3)}_{conf}=0.23 T\quad(0.1T)\eqno(26)$$
Here the number in brackets corresponds to a ``lowest order
approximation''
where $b_3$ is put to zero in the $\beta$-function (21).

For the symmetric phase of the electroweak theory we can neglect
the scalar fluctuations in a good approximation and the estimate (26)
directly applies. One has to deal with a strongly interacting gauge
theory with typical nonperturbative mass scales only somewhat below the
temperature scale! Similar to QCD one expects that condensates like
$\langle F_{ij}F^{ij}\rangle$ play an important role$^{15,24}$.
More generally, the physics of the symmetric phase  corresponds to a
strongly
coupled $SU(2)$ Yang-Mills theory in three dimensions: The relevant
excitations are ``$W$-balls'' (similar to glue balls) and scalar bound
states. All ``particles'' are massive (except the ``photon'') and
the relevant mass scale is set by $\Lambda_{conf}^{(3)}\sim T$. Also the
values of all condensates are given by appropriate powers of the
temperature.
Since the temperature is the only scale available the energy density must
have the same $T$-dependence as for an ideal gas
$$\rho=cT^4\eqno(27)$$
Only the coefficient $c$ should be different from the value obtained by
counting the perturbative degrees of freedom\footnote*{A similar remark
also applies to high temperature QCD. We expect quantitative modifications
of early cosmology due to the difference between $c$ and the ideal gas
value.}. We expect that quarks and leptons form $SU(2)$ singlet bound
states similar to the mesons in QCD\footnote*{We use here a language
appropriate for the excitations of the three-dimensional Euclidean
theory. Interpretation in terms of relativistic particles has to be used
with care!}. A chiral condensate seems, however,
unlikely in the high temperature regime and we do not think that fermions
play any important role for the dynamics of the electroweak phase
transition. The ``photon'' (or rather the gauge boson associated to
weak hypercharge) decouples from the $W$-balls. Its effective high
temperature
coupling to fermion and scalar bound states is renormalized to a very
small
value. As for the phase with spontaneous symmetry breaking, the fermions
and ``photon'' can be neglected for the symmetric phase. We conclude
that the high temperature phase transition of the electroweak theory can
be described by an effective three-dimensional Yang-Mills-Higgs system.
It is strongly interacting in the symmetric phase. Depending on the value
of the mass of the Higgs boson it may also be strongly interacting  in
the phase with spontaneous symmetry breaking if the temperature is near
the critical temperature. A more detailed investigation of this issue will
be given in the next section.

\vskip1cm\noindent
{\bf 5. RENORMALIZATION GROUP-IMPROVED EFFECTIVE POTENTIAL FOR THE
ELECTROWEAK PHASE TRANSITION}
\bigskip
Before performing a complete study of the average action for the standard
model at high temperature, the most important issues can be addressed in a
simplified treatment. For large enough $\rho,\rho\geq
\varphi^2_T,\varphi^2_T=8\pi^2 T^2/g^2_4$ (see sect. 2) we believe high
temperature perturbation theory and take the one-loop expression$^{6,7}$
$$
U_3(\rho_3)=-\mu^2(T)\rho_3+{1 \over 2}\left(\bar\lambda_3+{9\bar g_3^4
\over
128\pi^2T}\right)\rho^2_3
-{1 \over 12\pi}\left(6m^3_B+3m^3_E+m^3_1+3m^3_2\right)\eqno(28)$$
with
$$\eqalign{
m^2_B&={1 \over 2}\bar g^2_3\rho_3\cr
m^2_E&=m^2_D(T)+{1 \over 2}\bar g^2_3\rho_3\cr
m^2_1&=3\bar\lambda_3\rho_3-\mu^2(T)\cr
m^2_2&=\bar\lambda_3\rho_3-\mu^2(T)\cr}\eqno(29)$$
Here we use already a three-dimensional language with
$U_3=U/T,\rho_3=\rho/T,\bar\lambda_3=\lambda_4T,\bar g^2_3=g^2_4T$. The
couplings $\lambda_4$ and $g^2_4$ are evaluated at the scale
$k_T=2\pi T$ and $\mu^2(T),m^2_D(T)$ take the perturbative values. For
$\rho<\varphi_T^2$ we can compute a renormalization group-improved
potential by using a flow equation similar to (5) with $k\equiv m_B$.
In the language of the average action this amounts to an infrared cutoff
$P(x)=x+k^2$. This yields in the $\alpha=0$ gauge the approximative
evolution
equation$^{15}$
$$
{\partial \over \partial t}\bar\lambda_3(k)={3 \over 64\pi}
k^{-1}\Bigl\lbrace
2\bar g^4_3(k)+\bar g^4_3(k){m_B \over m_E}
+12\bar\lambda^2_3(k){m_B \over m_1}+4\bar\lambda^2_3(k){m_B \over m_2}
\bigr\rbrace\eqno(30)$$
We recover eq. (5) if we neglect $\bar\lambda_3$ on the r.h.s. and
approximate $m_B=m_E$. For an alternative derivation we could obtain
eq. (30) by taking appropriate derivatives of the potential $U_3(\rho_3)$
(28)
using the definitions (6) and treating the rations $m_B/m_E$ etc. as
independent of $k$.
For the correct interpretation of the flow equation we should also adapt
the relation between $k$ and $\rho$ to the situation with running gauge
coupling
$$k^2=m^2_B={1 \over 2}\bar g^2_3(k)\rho_3\eqno(31)$$
$$\bar\lambda_3={\partial^2U_3 \over \partial \rho^2_3}\eqno(32)$$
The solution of the flow equation (30) with initial values at $k_T$
specified
by (28) and the subsequent solution of the second order differential
equation
(32) for $U_3(\rho_3)$ should give a good approximation for the
temperature
dependent effective potential in a range of $\rho$ to be discussed below.

In eq. (30) the factor $m_B/m_E$ accounts properly for the Debye screening
of the $A_0$ mode. We take here $m_B/m_E=0$ for simplification of the
discussion.
We also observe that our equation becomes invalid for vanishing or
negative
$\bar\lambda\rho_3-\mu^2(T)$. This problem is related to the approach to
convexity of the effective scalar potential and absent in a more complete
treatment$^{25}$. We simplify by taking $\mu^2(T)=0$ in the mass ratios
such that $m^2_B/m^2_1={1 \over 6}\bar g^2_3/\bar\lambda_3, m^2_B/m^2_2={1
\over 2}
\bar g^2_3/\bar\lambda_3$ and
$${\partial \over \partial t}\bar\lambda_3={3 \over {32\pi}k}\left\lbrace
\bar g^4_3+(\sqrt 6+\sqrt 2)\bar\lambda_3^{3 \over 2}\bar g_3\right\rbrace
+2\eta\bar\lambda_3\eqno(33)$$
Here we have included the effects of wave function renormalization and
switched to a renormalized coupling $\bar\lambda_3$. The dominant
contribution
to the anomalous dimension $\eta$ comes from gauge boson fluctuations and
reads$^{15}$
$$\eta=-{1 \over {4\pi}}\bar g^2_3k^{-1}\eqno(34)$$
For the three-dimensional running of the gauge coupling we use eq. (21)
with
$b_3=0$ and
a correction factor $\tau$ accounting for the difference between the
$\alpha=0$ gauge needed here and the $\alpha=1$ gauge for which (21) was
computed
and for the contribution from scalar loops
$${\partial \over \partial t}\bar g^2_3=-{ 23 \over 24\pi}\tau\bar g^4_3
k^{-1}
\eqno(35)$$
We expect a value of $\tau$ near one. We can only use the system of
equations
(33) and (35) for $k$ larger than the three-dimensional confinement scale.
This gives a lower bound$^{15}$  on $\rho,\rho>\rho_{np}$:
Let us assume that renormalization group-improved perturbation theory
breaks down at a scale
$$k_{np}=1.1\Lambda^{(3)}_{conf}=1.1T{23\tau\alpha_w \over 6+{23\tau \over
2\pi}\alpha_w}\approx0.14 \tau T\eqno(36)$$
At this scale $\bar g^2_3(k_{np})$ has increased by a factor of about ten
as compared to $\bar g^2_3(k_T)$
$${\bar g_3^2(k_{np}) \over \bar g^2_3(k_T)}={11 \over 1+{23 \tau \alpha_W
\over 12\pi}}\approx 10.8\eqno(37)$$
and the dimensionless gauge coupling $g^2(k)=\bar g^2_3(k)/k$ has reached
the nonperturbative region
$${g^2(k_{np}) \over 4\pi^2}={60 \over 23\tau\pi}
=0.83\tau^{-1}\eqno(38)$$
{}From (31)(36) and (37) we infer
$$\rho_{np}=0.26{\alpha_w\tau^2 \over 1+{23\tau \alpha_W \over 12\pi}}
T^2\approx (0.09\tau T)^2\eqno(39)$$
For values of $\rho$ below $\rho_{np}$ we expect a complete breakdown
even for renormalization group-improved perturbation theory due
to condensates and similar strong interaction phenomena. We also observe
that the nonperturbative improvement of the evolution equation (21) for
$\bar g^2$ enhances $\rho_{np}$ by a factor of about 4.8. Our best
estimate
for the onset of strong nonperturbative phenomena is
therefore\footnote*{This
does not mean that high temperature perturbation theory can be trusted
for $\rho>\rho_{np}$. Only the renormalization group-improved potential
discussed here has a chance to remain valid at such low values of $\rho$!}
$$\rho_{np}=\left({T \over 5}\right)^2\eqno(40)$$

For $\rho>\rho_{np}$ it is instructive to consider the evolution of the
ratio
$\bar\lambda_3/\bar g^2_3$ and use $\bar g_3^2(k)$ instead of $k$ as a
variable
$${\partial \over \partial \ln \bar g^2_3}\left({\bar\lambda_3 \over
\bar g_3^2}\right)=-{9 \over 92\tau}\left\lbrace 1+{92\tau-48 \over
9}{\bar\lambda_3 \over \bar g^2_3}+(\sqrt 6+\sqrt 2)\left({\bar\lambda_3
\over
\bar g^2_3}\right)^{3/2}\right\rbrace\eqno(41)$$
This corresponds to the $\beta$-function for $\lambda/g^2$ shown in fig.
4.
The r.h.s. is  negative and $\bar\lambda_3$ is always driven from positive
towards negative values as $\bar g_3^2$ increases. The initial
ratio $\bar\lambda_3/\bar g^2_3$ at $k_T$ corresponds to the zero
temperature
ratio between Higgs scalar mass and $W$-boson mass up to small
(four-dimensional) logarithmic corrections\footnote{**}{This statement is
not valid for very small scalar masses when Coleman-Weinberg symmetry
breaking
operates.}
$${\bar\lambda_3 \over \bar g^2_3}(k_T)\approx\left({m_H \over
2m_W}\right)^2
\eqno(42)$$
For not too large values of $m_H/m_W$ we may, for the purpose of a
simplified
analytical discussion, neglect the last term in eq. (41) and use the
approximative solution
$$\eqalign{
&\bar\lambda_3(k)=\left({\bar g^2(k) \over \bar g^2(k_T)}\right)^{12 \over
23\tau}
\bar\lambda_3(k_T)\cr
&-{9 \over 92\tau-48}\bar g^2(k)\left(1-\left({\bar g^2(k_T) \over \bar
g^2(k)}\right)^{1-{12 \over 23\tau}}\right)\cr}\eqno(43)$$
The scale $k_{cw}$ (where $\bar\lambda_3(k_{cw})=0$) can be taken as a
characteristic scale for the phase transition. It obeys
$${\bar g^2(k_{cw}) \over \bar g^2(k_T)}=\left(1+{92\tau-48 \over 9}{\bar
\lambda_3(k_T) \over \bar g^2_3(k_T)}\right)^{23\tau \over 23\tau-12}
\eqno(44)$$
Equating $k_{cw}$ with $k_{np}$ (36) we find a critical value for the
ratio
$\bar\lambda_3(k_T)/\bar g^2_3(k_T)$ (for $\tau={24 \over 23}$)
$$\left({\bar\lambda_3(k_T) \over \bar
g^2_3(k_T)}\right)_{cr}=0.43\eqno(45)$$
We conclude that for a Higgs boson mass exceeding the critical value
$$(m_H)_{cr}\approx 1.3 m_W\approx 105\ {\rm GeV}\eqno(46)$$
the electroweak phase transition is described by a strongly interacting
$SU(2)$ gauge theory. Not only the symmetric phase but all phenomena
related to the phase transition are dominated by nonperturbative effects!
We emphasize that the critical value (46) should not be interpreted as an
accurate bound. Even for Higgs masses smaller than 100 GeV the strong
nonperturbative effects are very important and may dominate, for example,
the whole region of the effective potential between the origin and the
local maximum\footnote*{If we use the improved evolution equation (21)
instead
of (35), the critical value is lowered further.}. Our main conclusion is
that for
a Higgs boson mass of the order $m_W$ or even somewhat below it is
impossible to give a quantitative description of the phase transition
without taking
the strong  nonperturbative effects such as condensates into account.

One last remark concerns the ``nonperturbative region'' in the effective
potential for $\rho<\rho_{np}$ (39). (For $m_H$ taking the critical value
this concerns the region inside the turning point $\partial^2
U/\partial\rho^2
=0$.) Even though the relatively simple renormalization group-improved
treatment proposed in this section is not valid here, we can estimate the
difference
$\Delta U_3=U_3(\rho_{np})-U_3(0)$ by a simple scale argument: It has to
be proportional to the third power of $k_{np}$
$$\Delta U_3=K k^3_{np}\eqno(47)$$
Since there is no small dimensionless quantity in the problem and $k_{np}$
is determined relatively accurately, the constant $K$ should be near one.
Restoring the four-dimensional language we therefore estimate the
nonperturbative contribution to be roughly
$$\Delta U\approx 3\cdot 10^{-3} T^4\eqno(48)$$
We can also offer a speculative picture how the transition could
be described as an analytical crossover for very large
$m_H$: As the temperature raises a condensate $\langle
F_{ij}F^{ij}\rangle$
(or some other condensate)
may start forming at some
temperature $\tilde T$ for which the absolute minimum of the effective
scalar potential still occurs at $\rho_0(\tilde T)
\not=0$. For a further increase of the temperature beyond $\tilde T$ the
magnetic condensate  $\langle F_{ij}F^{ij}\rangle$ will increase whereas
$\rho_0(T)$ decreases. In the two-dimensional plane spanned by the
condensate
and $\rho$ the arrow ($\langle F_{ij}F^{ij}\rangle,\rho_0)$
may turn continuously from the $\rho$-direction for $T=\tilde T$ to the
condensate direction for very large temperatures. No jump in the particle
masses or other quantities would be expected for such a crossover. This
picture
gives a hint that a ``strongly interacting electroweak transition'' may
need
more degrees of freedom than the Higgs scalar for a
meaningful description of the vacuum structure!

We conclude that for realistic values of the Higgs boson mass
nonperturbative
techniques are necessary for a reliable description of the electroweak
phase transition. The ``strongly interacting electroweak phase
transition''
is in several aspects close to the high temperature phase transition in
QCD.
Similar methods for a description of both phenomena have to be developed,
in particular for an understanding of the temperature dependence of
various
condensates. This constitutes an interesting theoretical laboratory,
with possible applications ranging from early cosmology to high energy
heavy ion collisions.
\vskip1cm
\noindent{\bf Acknowledgement}
\bigskip
I am grateful to B. Bergerhoff, F. Freire,
D. Litim, S. Lola, M. Reuter, and N. Tetradis for collaboration on
high temperature gauge theories and to W. Buchm\"uller and M. Shaposhnikov
for fruitful discussions. I thank the organizers of the Sintra workshop
for a wonderful meeting.

\vskip1cm\noindent
{\bf REFERENCES}
\bigskip
\item{1.} D. A. Kirzhnitz, JETP Lett. {\bf 15}, 529 (1972);\hfill\break
D. A. Kirzhnits and A. D. Linde, Phys. Lett. {\bf B72}, 471 (1972).
\item{2.} V. A. Kuzmin, V. A. Rubakov, and M. E. Shaposhnikov,
Phys. Lett. {\bf B155}, 36 (1985);\hfill\break
M. E. Shaposhnikov, Nucl. Phys. {\bf B287}, 757 (1987); ibid. {\bf 299},
797
(1988).
\item{3.} A. D. Sakharov, JETP Lett. {\bf 5}, 24 (1967).
\item{4.} A. G. Cohen, D. B. Kaplan, and A. E. Nelson, Phys. Lett. {\bf
B245},
561 (1990); Nucl. Phys. {\bf B349} 727 (1991); \hfill\break
N. Turok and J. Zadrozny,
Phys. Rev. Lett. {\bf 65}, 2331 (1990); Nucl. Phys. {\bf B358}, 471
(1991);
\hfill\break
M. Dine, P. Huet, and R. Singleton, Nucl. Phys. {\bf B375}, 625 (1992);
\hfill\break
M. Dine, R. G. Leigh, P. Huet, A. D. Linde, and D. A. Linde, Phys. Rev.
{\bf D46}, 550 (1992);\hfill\break
G. R. Farrar and M. E. Shaposhnikov, preprint CERN-TH-6734-93;\hfill\break
M. B. Gavela, M. Lozano, J. Orloff, and O. P\`ene, preprints CERN-TH
7262/94,
7263/94.
\item{5.} A. I. Bochkarev and M. E. Shaposhnikov, Mod. Phys. Lett. {\bf
A2}, 417 (1987);\hfill\break
A. I. Bochkarev, S. V. Kuzmin, and M. E. Shaposhnikov, Phys. Lett. {\bf
B244},
257 (1990).
\item{6.} D. A. Kirzhnitz and A. D. Linde, Phys. Lett. {\bf 72B}, 471
(1972); JETP {\bf 40}, 628 (1974); Ann. Phys. {\bf 101}, 195
(1976);\hfill\break
S. Weinberg, Phys. Rev. {\bf D9}, 3357 (1974);\hfill\break
A. D. Linde, Nucl. Phys. {\bf B216}, 421 (1983), Rep. Prog. Phys.
{\bf 47}, 925 (1984).
\item{7.} L. Dolan and R. Jackiw, Phys. Rev. {\bf D9}, 3320 (1974);
\hfill\break
G. W. Anderson and L. J. Hall, Phys. Rev. {\bf D45}, 2685
(1992);\hfill\break
M. Carrington, Phys. Rev. {\bf D45}, 2933 (1992);\hfill\break
W. Buchm\"uller, Z. Fodor, T. Helbig, and D. Walliser, preprint
DESY 93-121, to appear in Ann. Phys.;\hfill\break
D. B\"odeker, W. Buchm\"uller, Z. Fodor, and T. Helbig, preprint
DESY 93-147, to appear in Nucl. Phys.;\hfill\break
J. R. Espinosa, M. Quiros, and F. Zwirner, Phys. Lett. {\bf B314}, 206
(1993).
\item{8.} N. Tetradis and C. Wetterich, Nucl. Phys. {\bf B398}, 659
(1993);
preprint HD-THEP-93-36.
\item{9.} T. Appelquist and R. Pisarski, Phys. Rev. {\bf D23}, 2305
(1981);
\hfill\break
S. Nadkarni, Phys. Rev. {\bf D27}, 917 (1983);\hfill\break
N. P. Landsman, Nucl. Phys. {\bf B322}, 498 (1989).
\item{10.} J. Kapusta, Finite temperature field theory, Cambridge
University
Press (1989).
\item{11.} T. Kaluza, Sitzungsber. Preuss. Akad. Wiss. Berlin, Math. Phys.
{\bf K1}, 966 (1921);\hfill\break
O. Klein, Z. f. Phys. {\bf 37}, 895 (1926).
\item{12.} D. O'Connor and C. R. Stephens, Nucl. Phys. {\bf B360}, 297
(1991),
preprint DIAS-STP-93-19;\hfill\break
D. O'Connor, C. R. Stephens, and F. Freire, Mod. Phys. Lett.
{\bf A8}, 1779 (1993), preprint THU 92/37.
\item{13.} A. Jakov\'ac, K. Kajantie, and A. Patk\'os, Helsinki Preprint
HU-TFT-94-01, hepph-9312355;\hfill\break
K. Farakos, K. Kajantie, K. Rummukainen, and M. Shaposhnikov, preprint
CERN-TH6973/94.
\item{14.} P. Lacock, D. E. Miller, and T. Reisz, Nucl. Phys. {\bf B369},
501 (1992);\hfill\break
L. K\"arkk\"ainen, P. Lacock, B. Petersson, and T. Reisz, Nucl. Phys.
{\bf B395}, 733 (1993).
\item{15.} M. Reuter, C. Wetterich, Nucl. Phys. {\bf B391}, 147 (1993),
and Nucl. Phys. {\bf B408}, 91 (1993); Heidelberg preprints
HD-THEP-93-40/41,
to appear in Nucl. Phys. B.
\item{16.} S. Coleman and E. Weinberg, Phys. Rev. {\bf D7}, 1888 (1973).
\item{17.} A. D. Linde, Phys. Lett. {\bf 96B}, 289 (1980);\hfill\break
D. Gross, R. Pisarski, and L. Yaffe, Rev. Mod. Phys. {\bf 53}, 43 (1981).
\item{18.} C. Wetterich, Nucl. Phys. {\bf B352}, 529 (1991).
\item{19.} L. P. Kadanoff, Physica {\bf 2}, 263 (1966);\hfill\break
K. G. Wilson, Phys. Rev. {\bf B4}, 3184 (1971);\hfill\break
K. G. Wilson, I. G. Kogut, Phys. Rep. {\bf 12}, 75 (1974);\hfill\break
F. Wegner in: {\it Phase transitions and critical phenomena}, vol. 6, eds.
C. Domb and M. S. Green (Academic Press, New York 1976);\hfill\break
G. Mack. T. Kalkreuter, G. Palma, M. Speh, Int. J. Mod. Phys. C, Vol. 3,
No. 1,
121-47 (1992).
\item{20.} C. Wetterich, Phys. Lett. {\bf B301}, 90 (1993);\hfill\break
C. Wetterich, Z. Phys. {\bf C60}, 461 (1993).
\item{21.} F. Wegner, A. Houghton, Phys. Rev. {\bf A8}, 401
(1973);\hfill\break
K. G. Wilson, I. G. Kogut, Phys. Rep. {\bf 12}, 75 (1974);\hfill\break
S. Weinberg, Critical Phenomena for Field Theorists, Erice Subnucl. Phys.
1
(1976);\hfill\break
J. Polchinski, Nucl. Phys. {\bf B231}, 269 (1984);\hfill\break
A. Hasenfratz, P. Hasenfratz, Nucl. Phys. {\bf B270}, 685 (1986).
\item{22.} F. J. Dyson, Phys. Rev. {\bf 75}, 1736 (1949); \hfill\break
J. Schwinger, Proc. Nat. Acad. Sc. {\bf 37}, 452, 455 (1951).
\item{23.} P. Ginsparg, Nucl. Phys. {\bf B170} [FS1], 388
(1980);\hfill\break
P. Arnold and L. Yaffe, Preprint UW/PT-93-24 (1993);\hfill\break
W. Buchm\"uller and Z. Fodor, preprint DESY 94-045.
\item{24.} M. Shaposhnikov, Phys. Lett. {\bf B316}, 112 (1993).
\item{25.} A. Ringwald and C. Wetterich, Nucl. Phys. {\bf B334}, 506
(1990);\hfill\break
N. Tetradis and C. Wetterich, Nucl. Phys. {\bf B383}, 197 (1992).
\end